\documentclass[preprint,12pt,authoryear]{elsarticle}

\usepackage{amssymb}

\journal{Nucl. Instr. and Meth. in Phys. Res. A}

\begin{document}

\begin{frontmatter}



\title{On Ruby's solid angle formula and some of its generalizations}


\author{Samuel Friot\footnote{Part of this work has been done at Institut de Physique Nucl\'eaire de Lyon, Universit\'e Lyon 1, IN2P3-CNRS, F-69622 Villeurbanne Cedex, France.}}

\address{Institut de Physique Nucl\' eaire d'Orsay

Universit\' e Paris-Sud 11, IN2P3-CNRS, F-91405 Orsay Cedex, France}

\begin{abstract}
Using the Mellin-Barnes representation, we show that Ruby's solid angle formula and some of its generalizations may be expressed in a compact way in terms of the Appell $F_4$ and Lauricella $F_C$ functions.

\end{abstract}


\end{frontmatter}


\section{Introduction}
\label{}

Ruby's formula, giving the solid angle subtended at a disk source by a coaxial parallel-disk detector  [\cite{Ruby}], is the following:
\begin{equation}\label{ruby}
G=\frac{R_D}{R_S}\int_0^\infty dk \frac{e^{-kd}}{k}J_1(kR_S)J_1(kR_D),
\end{equation}
where  $R_S$ and $R_D$ are respectively the radius of the source and of the detector, $d$ is the distance between the source and the detector and $J_1(x)$ is the Bessel function of first kind and order 1.

Until [\cite{Conway}], where an expression in terms of complete and incomplete elliptic integrals has been given, it seems that Ruby's formula had not been expressed in a closed form. A double series representation had been previously mentioned in [\cite{Ruby2}] but it was concluded in [\cite{Pomme}] that the convergence region of this double series is restricted in a way that when the detector is too close to the source, one had to compute the integral by means of numerical methods.

In the next section, we will see that with the help of the Mellin-Barnes (MB) representation method (see e.g. [\cite{Kaminski}] for an introduction), Ruby's formula may be expressed in a compact way in terms of the Appell function $F_4$.

Generalizations of Ruby's formula, which have been treated mainly in [\cite{Conway}] but not always obtained in closed form, will also be considered within the same approach, in a subsequent section where their expressions in terms of the Lauricella function $F_C$ will be given.

\section{Ruby's formula}

The Bessel function $J_1(z)$ has the following MB representation, valid for $z>0$, see [\cite{Kaminski}]:
\begin{equation}\label{MB_bessel}
J_1(z)=\frac{1}{2i\pi}\int_{c-i\infty}^{c+i\infty}ds\left(\frac{z}{2}\right)^{1-2s}\frac{\Gamma(s)}{\Gamma(2-s)},
\end{equation}
where, for absolute convergence of the integral, the constant $c$, which is the real part of $s$ (since the chosen integration path is a vertical line in the $s$-complex plane) has to belong to the interval $\left]0,\frac{1}{2}\right[$ [\cite{Kaminski}].

Inserting twice this integral representation in eq. (\ref{ruby}) we get 
\begin{eqnarray}\nonumber
G&=&\frac{R_D^2}{4}\left(\frac{1}{2i\pi}\right)^2\! \int_{c-i\infty}^{c+i\infty}ds\int_{c'-i\infty}^{c'+i\infty}dt\left(\frac{R_S}{2}\right)^{-2s}\!\left(\frac{R_D}{2}\right)^{-2t}\frac{\Gamma(s)}{\Gamma(2-s)}\frac{\Gamma(t)}{\Gamma(2-t)}\\
&\times & \! \int_0^\infty dk\ k^{1-2s-2t}e^{-kd},
\end{eqnarray}
where $c=\Re(s)\in\left]0,\frac{1}{2}\right[$ and $c'=\Re(t)\in\left]0,\frac{1}{2}\right[$.

Performing the  $k$-integral leads to the following 2-fold MB representation: 
\begin{eqnarray}
G&=&\left(\frac{1}{2i\pi}\right)^2\left(\frac{R_D}{2d}\right)^2\\
&\times &\! \int_{c-i\infty}^{c+i\infty}ds\int_{c'-i\infty}^{c'+i\infty}dt\left(\frac{R_S}{2d}\right)^{-2s}\!\left(\frac{R_D}{2d}\right)^{-2t}\frac{\Gamma(s)\Gamma(t)\Gamma(2-2s-2t)}{\Gamma(2-s)\Gamma(2-t)},\nonumber
\end{eqnarray}
with the constraint $\Re(s+t)<1$, which is fulfilled.

To compute this integral one can directly apply the general method described in [\cite{Friot}]. If one follows this approach, one will find three double series representations (one of them being the one mentioned in [\cite{Ruby2}] and studied in [\cite{Pomme}]), converging in three different regions of values of the parameters $R_S$, $R_D$ and $d$.

It is however even simpler to notice that, by using the duplication formula for the Euler gamma function
\begin{equation}\label{duplication}
\Gamma(2s)=\frac{1}{\sqrt{\pi}}2^{2s-1}\Gamma(s)\Gamma\left(s+\frac{1}{2}\right),
\end{equation}
one has
\begin{eqnarray}\label{MB_F4}
G&=&\frac{2}{\sqrt \pi}\left(\frac{R_D}{2d}\right)^2\left(\frac{1}{2i\pi}\right)^2\int_{c-i\infty}^{c+i\infty}ds\int_{c'-i\infty}^{c'+i\infty}dt\left(\frac{R_S}{d}\right)^{-2s}\!\left(\frac{R_D}{d}\right)^{-2t}\nonumber\\
&\times &\! \frac{\Gamma(s)\Gamma(t)}{\Gamma(2-s)\Gamma(2-t)}\Gamma(1-s-t)\Gamma\left(\frac{3}{2}-s-t\right),
\end{eqnarray}
which, since $\Gamma\left(\frac{3}{2}\right)=\frac{2}{\sqrt \pi}$, is nothing but the MB representation of the Appell $F_4$ function [\cite{Appell}] 
\begin{equation}\label{F4ruby}
G=\left(\frac{R_D}{2d}\right)^2F_4\left(1,\frac{3}{2},2,2;-\left(\frac{R_S}{d}\right)^2,-\left(\frac{R_D}{d}\right)^2\right)
\end{equation}
from where, by definition, one gets the double series representation
\begin{equation}\label{F4series}
G=\left(\frac{R_D}{2d}\right)^2\sum_{m=0}^\infty\sum_{n=0}^\infty\frac{(-1)^{m+n}}{m!n!}\frac{(1)_{m+n}\left(\frac{3}{2}\right)_{m+n}}{(2)_m(2)_n}\left(\frac{R_S}{d}\right)^{2m}\left(\frac{R_D}{d}\right)^{2n},
\end{equation}
where $(a)_m=\frac{\Gamma(a+m)}{\Gamma(a)}$ is the Pochhammer symbol.

The series (\ref{F4series}) is the same as the one studied in [\cite{Pomme}] with convergence region given\footnote{The convergence regions of the double series representations of the Appell functions are well-known: it is straightforward to get them by Horn's method, see for instance [\cite{Friot}] or [\cite{Appell}].} by $R_S+R_D<d$. This confirms the analysis performed in [\cite{Pomme}].

The well-known analytic continuation formula [\cite{Appell}]
\begin{eqnarray}\nonumber\label{analytic_continuation}
&F_4&\!\!\!\!\!\left(a,b,c,d;x,y\right)=\frac{\Gamma(d)\Gamma(b-a)}{\Gamma(d-a)\Gamma(b)}(-y)^{-a}F_4\left(a,a+1-d,c,a+1-b;\frac{x}{y},\frac{1}{y}\right)\\
&+&\frac{\Gamma(d)\Gamma(a-b)}{\Gamma(d-b)\Gamma(a)}(-y)^{-b}F_4\left(b+1-d,b,c,b+1-a;\frac{x}{y},\frac{1}{y}\right)
\end{eqnarray}
and the symmetric relation obtained by exchanging $x$ with $y$ and $c$ with $d$
allow to obtain double series representations valid in the regions $R_S+d<R_D$ and $R_D+d<R_S$.

Series representations valid in other ranges of values of the parameters than those given above may be found in [\cite{Exton}], where a full analytic continuation study has been performed.

\section{Generalization}

The same technique may be used to compute the more general integral
\begin{equation}\label{conway_gen}
I_{(l,m_1,...,m_N)}=\int_0^\infty dk\ k^l\ e^{-kd}\prod_{j=1}^{N}J_{m_j}(kR_j),
\end{equation}
where $l$ and the $m_j$ are such that the integral converges, and $R_j>0$ for all $j\in\{1,...,N\}$.

In this case, we use the following MB representation for the Bessel functions (valid for $z>0$):
\begin{equation}\label{MB_bessel_gen}
J_m(z)=\frac{1}{2i\pi}\int_{c-i\infty}^{c+i\infty}ds\left(\frac{z}{2}\right)^{m-2s}\frac{\Gamma(s)}{\Gamma(1+m-s)},
\end{equation}
where $c=\Re(s)\in\left]0,\frac{\Re(m)}{2}\right[$. 

Notice that this integral is defined only when $m>0$ but we will see that one can relax this constraint at the end of the calculations by appeal to analytic continuation.

Inserting eq. (\ref{MB_bessel_gen}) in eq. (\ref{conway_gen}) we get
\begin{eqnarray}\nonumber
I_{(l,m_1,...,m_N)}&=&\prod_{j=1}^{N}\left[\frac{1}{2i\pi}\int_{c_j-i\infty}^{c_j+i\infty}ds_j\left(\frac{R_j}{2}\right)^{m_j-2s_j}\frac{\Gamma(s_j)}{\Gamma(1+m_j-s_j)}\right]\\
&\times & \! \int_0^\infty dk\ e^{-kd}k^{l+\sum_{j=1}^N(m_j-2s_j)}.
\end{eqnarray}
The $k$-integral gives $d^{-1-l-\sum_{j=1}^N(m_j-2s_j)}\Gamma\left(1+l+\sum_{j=1}^N(m_j-2s_j)\right)$ with the constraint $\Re\left(1+l+\sum_{j=1}^N(m_j-2s_j)\right)>0$. Let us suppose that this constraint is fulfilled (in all particular cases considered in [\cite{Conway}], it is always possible to satisfy this constraint by an appropriate choice of the $c_j$).

Then, applying eq. (\ref{duplication}), one may conclude that  
\begin{eqnarray}\nonumber\label{MB_general}
&I&_{\!\!\!\!\!\!\!\!\!(l,m_1,...,m_N)}\\ \nonumber
&=&\frac{1}{\sqrt{\pi}}\left(\frac{2}{d}\right)^l\frac{1}{d}\prod_{j=1}^{N}\left[\left(\frac{R_j}{d}\right)^{m_j}\frac{1}{2i\pi}\int_{c_j-i\infty}^{c_j+i\infty}ds_j\left(\frac{R_j}{d}\right)^{-2s_j}\frac{\Gamma(s_j)}{\Gamma(1+m_j-s_j)}\right]\\
&\times & \! \Gamma\left(\sum_{j=1}^N\left(\frac{m_j}{2}-s_j\right)+\frac{l+1}{2}\right)\Gamma\left(\sum_{j=1}^N\left(\frac{m_j}{2}-s_j\right)+\frac{l}{2}+1\right).
\end{eqnarray}
As a particular check, it is easy to derive eq. (\ref{MB_F4}) from eq. (\ref{MB_general}) by putting $N=2$, $m_1=m_2=1$, $R_1=R_D$, $R_2=R_S$ and $l=-1$, and multiplying by $\frac{R_D}{R_S}$.

In the case where $\sum_{j=1}^N\frac{m_j}{2}+\frac{l+1}{2}$ and $\sum_{j=1}^N\frac{m_j}{2}+\frac{l}{2}+1$ are positive numbers, we recognize in eq. (\ref{MB_general}) the MB representation of the multiple Lauricella function $F_C^{(N)}$ (modulo an overall factor) [\cite{Exton2}]. 

We therefore have
\begin{eqnarray}\label{lauricella}
&I&_{\!\!\!\!\!\!\!\!\!(l,m_1,...,m_N)}\\ \nonumber
&=&\frac{1}{\sqrt{\pi}}\left(\frac{2}{d}\right)^l\frac{1}{d}\frac{\Gamma\left(\sum_{j=1}^N\frac{m_j}{2}+\frac{l+1}{2}\right)\Gamma\left(\sum_{j=1}^N\frac{m_j}{2}+\frac{l}{2}+1\right)}{\prod_{j=1}^{N}\Gamma(1+m_j)}\prod_{j=1}^{N}\left(\frac{R_j}{d}\right)^{m_j}\\ \nonumber
&&\!\!\!\!\!\!\!\!\!\!\!\!\!\!\!\!\times F_C^{(N)}\left(\sum_{j=1}^N\frac{m_j}{2}+\frac{l+1}{2},\sum_{j=1}^N\frac{m_j}{2}+\frac{l}{2}+1,1+m_1,...,1+m_N;-\frac{R_1^2}{d^2},...,-\frac{R_N^2}{d^2}\right)\!\! .
\end{eqnarray}
Lauricella functions are the generalizations of Appell functions and $F_C^{(2)}$ is, obviously, nothing but the Appell $F_4$ function.

The integral representation in eq. (\ref{MB_general}) has been obtained with the initial constraint that the $m_j$ and $R_j$ are strictly positive for all $j\in\{1,...,N\}$. By analytic continuation, it is however possible to include other values for these parameters (and, among others, the important case where some of the $m_j$ are equal to zero) since eq. (\ref{MB_general}) is defined as long as the quantities $\sum_{j=1}^N\frac{m_j}{2}+\frac{l+1}{2}$ and $\sum_{j=1}^N\frac{m_j}{2}+\frac{l}{2}+1$ are not negative integers. In fact, in all particular situations considered in [\cite{Conway}] these two quantities are positive numbers. Therefore one may directly use eq. (\ref{lauricella}) to compute them and this is done in a subsection to follow.

Using the multiple series representation of the multiple Lauricella function $F_C^{(N)}$ [\cite{Exton2}], one obtains:
\begin{eqnarray}\label{FCseries}\nonumber
&I&_{\!\!\!\!\!\!\!\!\!(l,m_1,...,m_N)}\\ \nonumber
&=&\frac{1}{\sqrt{\pi}}\left(\frac{2}{d}\right)^l\frac{1}{d}\frac{\Gamma\left(\sum_{j=1}^N\frac{m_j}{2}+\frac{1+l}{2}\right)\Gamma\left(\sum_{j=1}^N\frac{m_j}{2}+\frac{l}{2}+1\right)}{\prod_{j=1}^{N}\Gamma(1+m_j)}\prod_{j=1}^{N}\left(\frac{R_j}{d}\right)^{m_j}\\ \nonumber
&&\times \sum_{k_1=0}^{\infty}...\sum_{k_N=0}^{\infty}\frac{\left(\sum_{j=1}^N\frac{m_j}{2}+\frac{l+1}{2}\right)_{\sum_{j=1}^N k_j}\left(\sum_{j=1}^N\frac{m_j}{2}+\frac{l}{2}+1\right)_{\sum_{j=1}^N k_j}}{\prod_{j=1}^{N}\left((1+m_j)_{k_j}k_j!\right)}\\
 &\times & (-1)^{\sum_{j=1}^N k_j}\left(\frac{R_1}{d}\right)^{2k_1}...\left(\frac{R_N}{d}\right)^{2k_N}
\end{eqnarray}
where as before $(a)_m$ is the Pochhammer symbol.

This multiple series converges in the region $\sum_{j=1}^N |R_j|<|d|$ [\cite{Exton2}].

An analytic continuation formula for $F_C^{(N)}$, generalizing eq. (\ref{analytic_continuation}), is [\cite{Exton2}]:
\begin{eqnarray}\nonumber\label{analytic_continuation_FC}
&F_C^{(N)}&\!\!\!\!\!\left(a,b,c_1,...,c_N;x_1,...,x_N\right)=\frac{\Gamma(c_N)\Gamma(b-a)}{\Gamma(c_N-a)\Gamma(b)}(-x_N)^{-a}\\ \nonumber
&\times & F_C^{(N)}\left(a,a+1-c_N,c_1,...,c_N,a+1-b;\frac{x_1}{x_N},...,\frac{x_{N-1}}{x_N},\frac{1}{x_N}\right)\\ \nonumber
&+&\frac{\Gamma(c_N)\Gamma(a-b)}{\Gamma(c_N-b)\Gamma(a)}(-x_N)^{-b}\\
&\times & F_C^{(N)}\left(b+1-c_N,b,c1,...,c_N,b+1-a;\frac{x_1}{x_N},...,\frac{x_{N-1}}{x_N},\frac{1}{x_N}\right).
\end{eqnarray}
From this equation, one can obtain a series representation valid in the region $\sqrt{\left|\frac{x_1}{x_N}\right|}+...+\sqrt{\left|\frac{1}{x_N}\right|}<1$. As in the case of $F_4$, symmetric expressions and their corresponding convergence regions are straightforward to obtain.

 \subsection{Particular cases}

To ease the use of previous equations, we give explicit expressions for the cases where $N=2$ and $N=3$. We suppose that the constraints discussed above are checked.
\begin{eqnarray}\nonumber
&I&_{\!\!\!\!\!\!\!\!\!(l,m_1,m_2)}=\int_0^\infty dk\ k^l\ e^{-kd}J_{m_1}(kR_1)J_{m_2}(kR_2)\\ \nonumber
&=&\frac{1}{\sqrt{\pi}}\left(\frac{2}{d}\right)^l\frac{1}{d}\left(\frac{R_1}{d}\right)^{m_1}\left(\frac{R_2}{d}\right)^{m_2}\left(\frac{1}{2i\pi}\right)^2\int_{c_1-i\infty}^{c_1+i\infty}\int_{c_2-i\infty}^{c_2+i\infty}ds_1ds_2\\ \nonumber
&\times &\left(\frac{R_1}{d}\right)^{-2s_1}\left(\frac{R_2}{d}\right)^{-2s_2}\frac{\Gamma(s_1)\Gamma(s_2)}{\Gamma(1+m_1-s_1)\Gamma(1+m_2-s_2)}\\ \nonumber
&\times & \! \Gamma\left(\frac{m_1}{2}+\frac{m_2}{2}-s_1-s_2+\frac{l+1}{2}\right)\Gamma\left(\frac{m_1}{2}+\frac{m_2}{2}-s_1-s_2+\frac{l}{2}+1\right)\\ \nonumber
&=&\frac{1}{\sqrt{\pi}}\left(\frac{2}{d}\right)^l\frac{1}{d}\left(\frac{R_1}{d}\right)^{m_1}\left(\frac{R_2}{d}\right)^{m_2}\frac{\Gamma\left(\frac{m_1}{2}+\frac{m_2}{2}+\frac{l+1}{2}\right)\Gamma\left(\frac{m_1}{2}+\frac{m_2}{2}+\frac{l}{2}+1\right)}{\Gamma(1+m_1)\Gamma(1+m_2)}\\ \nonumber
&\times & F_4\left(\frac{m_1}{2}+\frac{m_2}{2}+\frac{l+1}{2},\frac{m_1}{2}+\frac{m_2}{2}+\frac{l}{2}+1,1+m_1,1+m_2;-\frac{R_1^2}{d^2},-\frac{R_2^2}{d^2}\right)\\ \nonumber
&=&\frac{1}{\sqrt{\pi}}\left(\frac{2}{d}\right)^l\frac{1}{d}\left(\frac{R_1}{d}\right)^{m_1}\left(\frac{R_2}{d}\right)^{m_2}\\ \nonumber
&\times &\sum_{k_1=0}^{\infty}\sum_{k_2=0}^{\infty}\frac{\Gamma\left(\frac{m_1}{2}+\frac{m_2}{2}+\frac{l+1}{2}+k_1+k_2\right)\Gamma\left(\frac{m_1}{2}+\frac{m_2}{2}+\frac{l}{2}+1+k_1+k_2\right)}{\Gamma(1+m_1+k_1)\Gamma(1+m_2+k_2)k_1!k_2!}
\\ 
&\times & \left(-\frac{R_1^2}{d^2}\right)^{k_1}\left(-\frac{R_2^2}{d^2}\right)^{k_2}.
\end{eqnarray}

\begin{eqnarray}\nonumber
&I&_{\!\!\!\!\!\!\!\!\!(l,m_1,m_2,m_3)}=\int_0^\infty dk\ k^l\ e^{-kd}J_{m_1}(kR_1)J_{m_2}(kR_2)J_{m_3}(kR_3)\\ \nonumber
&=&\frac{1}{\sqrt{\pi}}\left(\frac{2}{d}\right)^l\frac{1}{d}\left(\frac{R_1}{d}\right)^{m_1}\left(\frac{R_2}{d}\right)^{m_2}\left(\frac{R_3}{d}\right)^{m_3}\\ \nonumber
&\times &\left(\frac{1}{2i\pi}\right)^3\int_{c_1-i\infty}^{c_1+i\infty}\int_{c_2-i\infty}^{c_2+i\infty}\int_{c_3-i\infty}^{c_3+i\infty}ds_1ds_2ds_3\left(\frac{R_1}{d}\right)^{-2s_1}\!\!\left(\frac{R_2}{d}\right)^{-2s_2}\!\!\left(\frac{R_3}{d}\right)^{-2s_3}\\ \nonumber
&\times &\frac{\Gamma(s_1)\Gamma(s_2)\Gamma(s_3)}{\Gamma(1+m_1-s_1)\Gamma(1+m_2-s_2)\Gamma(1+m_3-s_3)}\\ \nonumber
&\times & \! \Gamma\left(\frac{m_1}{2}+\frac{m_2}{2}+\frac{m_3}{2}-s_1-s_2-s_3+\frac{l+1}{2}\right)\\ \nonumber
&\times &\Gamma\left(\frac{m_1}{2}+\frac{m_2}{2}+\frac{m_3}{2}-s_1-s_2-s_3+\frac{l}{2}+1\right)\\ \nonumber
&=&\frac{1}{\sqrt{\pi}}\left(\frac{2}{d}\right)^l\frac{1}{d}\left(\frac{R_1}{d}\right)^{m_1}\left(\frac{R_2}{d}\right)^{m_2}\left(\frac{R_3}{d}\right)^{m_3}\\ \nonumber
&\times &\frac{\Gamma\left(\frac{m_1}{2}+\frac{m_2}{2}+\frac{m_3}{2}+\frac{l+1}{2}\right)\Gamma\left(\frac{m_1}{2}+\frac{m_2}{2}+\frac{m_3}{2}+\frac{l}{2}+1\right)}{\Gamma(1+m_1)\Gamma(1+m_2)\Gamma(1+m_3)}\\ \nonumber
&\times & F_C\left(\frac{m_1}{2}+\frac{m_2}{2}+\frac{m_3}{2}+\frac{l+1}{2},\frac{m_1}{2}+\frac{m_2}{2}+\frac{m_3}{2}+\frac{l}{2}+1,1+m_1,1+m_2,\right.\\  \label{3Bessel}
&&\left.1+m_3;-\frac{R_1^2}{d^2},-\frac{R_2^2}{d^2},-\frac{R_3^2}{d^2}\right)\\ \nonumber
&=&\frac{1}{\sqrt{\pi}}\left(\frac{2}{d}\right)^l\frac{1}{d}\left(\frac{R_1}{d}\right)^{m_1}\left(\frac{R_2}{d}\right)^{m_2}\left(\frac{R_3}{d}\right)^{m_3}\\ \nonumber
&\times &\sum_{k_1=0}^{\infty}\sum_{k_2=0}^{\infty}\sum_{k_3=0}^{\infty}\frac{\Gamma\left(\frac{m_1}{2}+\frac{m_2}{2}+\frac{m_3}{2}+\frac{l+1}{2}+k_1+k_2+k_3\right)}{\Gamma(1+m_1+k_1)\Gamma(1+m_2+k_2)\Gamma(1+m_3+k_3)k_1!k_2!k_3!}
\\ \nonumber
\times & \Gamma&\!\!\!\!\!\left(\frac{m_1}{2}+\frac{m_2}{2}+\frac{m_3}{2}+\frac{l}{2}+1+k_1+k_2+k_3\right)\left(-\frac{R_1^2}{d^2}\right)^{k_1}\!\!\left(-\frac{R_2^2}{d^2}\right)^{k_2}\!\!\left(-\frac{R_3^2}{d^2}\right)^{k_3}.
\end{eqnarray}

 \subsection{Applications}

We will now consider several particular cases studied in [\cite{Conway}].

First, the non-coaxial situation where the source disk has a constant surface emissivity. In this case
\begin{equation}\label{noncoax}
G=\frac{R_D}{R_S}\int_0^\infty dk \frac{e^{-kd}}{k}J_0(ka)J_1(kR_S)J_1(kR_D),
\end{equation}
where $a$ is the offset distance of the disks axes. 

It is clear from the discussion above that putting $N=3$, $l=-1$, $m_1=0$ and $m_2=m_3=1$  in eq. (\ref{lauricella}) (or in eq. (\ref{3Bessel})) we get
\begin{equation}\label{constant_emissivity}
G=\left(\frac{R_D}{2d}\right)^2F_C\left(1,\frac{3}{2},1,2,2;-\left(\frac{a}{d}\right)^2, -\left(\frac{R_S}{d}\right)^2, -\left(\frac{R_D}{d}\right)^2\right).
\end{equation}
The corresponding triple series representation, converging in the region $a+R_D+R_S<d$,  may be obtained from eq. (\ref{FCseries}). It matches with the triple series of [\cite{Pomme2}]. Other triple series, converging in other regions, may easily be found from eq. (\ref{analytic_continuation_FC}).

An example of non-constant surface emissivity (parabolic radial distribution) has also been considered in [\cite{Conway}], the corresponding formula being
\begin{equation}
G_1=\frac{4R_D}{R_S^2}\int_0^\infty dk \frac{e^{-kd}}{k^2}J_0(ka)J_2(kR_S)J_1(kR_D).
\end{equation}
This gives $N=3$, $l=-2$, $m_1=0$, $m_2=2$ and $m_3=1$ and then
\begin{equation}
G_1=\left(\frac{R_D}{2d}\right)^2F_C\left(1,\frac{3}{2},1,3,2;-\left(\frac{a}{d}\right)^2, -\left(\frac{R_S}{d}\right)^2, -\left(\frac{R_D}{d}\right)^2\right).
\end{equation}
Now, let us consider the non-coaxial ring source case, where
\begin{equation}
G=\frac{R_D}{2}\int_0^\infty dk\  e^{-kd}J_0(ka)J_0(kR_S)J_1(kR_D).
\end{equation}
In this case, we have $N=3$, $l=0$, $m_1=m_2=0$ and $m_3=1$.

This leads again to an expression very similar to eq. (\ref{constant_emissivity}):
\begin{equation}
G=\left(\frac{R_D}{2d}\right)^2F_C\left(1,\frac{3}{2},1,1,2;-\left(\frac{a}{d}\right)^2, -\left(\frac{R_S}{d}\right)^2, -\left(\frac{R_D}{d}\right)^2\right).
\end{equation}

\section{Conclusions}

In this note, we showed that by using the MB representation method, it has been possible to compute a family of integrals, see eq. (\ref{conway_gen}), which include Ruby's solid angle formula as a particular case.

Notice that integrals of this family involving three Bessel functions also appear in other contexts, see for instance [\cite{Conway2}, \cite{Conway3}] (and some of the references therein by the same author). In these references, it is claimed that these particular integrals have not yet been calculated in closed-form. By computing the whole family in this paper, we therefore fill this gap.

We also would like to add the following remark. We have been able to recognize the closed-form expression for Ruby's formula, eq. (\ref{F4ruby}), directly from the MB representation, eq. (\ref{MB_F4}). Then, analytical continuations could be easily obtained from the known analytical continuation formulas of the Appell $F_4$ function. 
However, let us imagine now that we would have been in a less simple case where we could not recognize any known function from the MB representation, or even from a double series representation (similar to eq. (\ref{F4series}), computed from the MB representation or by another method). The real power of the MB method rests on the fact that, for a large class of MB integrals, we still would have been able to obtain several series representations, analytic continuations of each other, following the method presented in [\cite{Friot}].





\end{document}